\documentclass{nature}
\usepackage{graphicx}
\makeatletter
\let\saved@includegraphics\includegraphics
\AtBeginDocument{\let\includegraphics\saved@includegraphics}
\renewenvironment*{figure}{\@float{figure}}{\end@float}
\makeatother
\usepackage[utf8]{inputenc}
\usepackage[T1]{fontenc}
\usepackage[UKenglish]{babel}
\usepackage{xcolor}
\usepackage{amsmath}
\usepackage{amssymb}
\usepackage{indentfirst}
\usepackage{fancyhdr}
\usepackage{ulem}

\title{Rotational Quantum Beat Lasing Without Inversion} 

\author{Maria Richter$^{1,\dagger,\ast}$, Marianna Lytova$^{2,\dagger}$, Felipe Morales$^{1}$, Stefan Haessler$^{3}$, Olga Smirnova$^{1,4}$, Michael Spanner$^{2,5}$, \& Misha Ivanov$^{1,6,7}$}

\begin{document}

\maketitle

\begin{affiliations}
 \item Max-Born-Institute, Max-Born Stra{\ss}e 2A, 12489 Berlin, Germany.
 \item Department of Physics, University of Ottawa, Ottawa K1N 6N5, Canada.
 \item Laboratoire d’Optique Appliquée, CNRS, École Polytechnique, ENSTA Paris, Institut Polytechnique de Paris, 181 Chemin de la Hunière et des Joncherettes, 91120 Palaiseau, France.
 \item Technische Universit\"at Berlin, Ernst-Ruska-Geb\"aude, Hardenbergstra{\ss}e 36A, 10623 Berlin, Germany.
 \item National Research Council of Canada, 100 Sussex Drive, Ottawa K1A 0R6, Canada.
 \item Department of Physics, Humboldt University, Newtonstra{\ss}e 15, D-12489 Berlin, Germany.
 \item Blackett Laboratory, Imperial College London, 
 SW7 2AZ London, United Kingdom.
\end{affiliations}
\noindent $\dagger$ These authors contributed equally,
\noindent $\ast$ Corresponding author

\begin{abstract}
In standard lasers, light amplification requires population inversion between an upper and a lower state to break the reciprocity between absorption and stimulated emission. However, in a medium prepared in a specific superposition state, quantum interference may fully suppress absorption while leaving stimulated emission intact, opening the possibility of lasing without inversion. 
Here we show that lasing without inversion arises naturally 
during  propagation of intense femtosecond laser pulses in air. 
It is triggered by the combination of molecular ionization and molecular alignment, both 
unavoidable in 
intense light fields. The effect 
could enable inversionless amplification of broadband
radiation in many molecular gases, opening unusual opportunities for remote sensing.  
\end{abstract}
\section*{Introduction}
A resonant light propagating through a medium will be absorbed by the medium in a lower state  and will
stimulate emission if the medium is excited. For a quantum system pumped into an excited state $|e\rangle$, with an empty state $|g\rangle$ below, population inversion 
between these two states leads to amplification of 
a probe light at the $|e\rangle \rightarrow |g\rangle$ transition frequency. 
For almost six decades, this process has been exploited in 
conventional lasers. However, already three decades ago, 
it has been shown that lasing can also occur  
without population inversion\cite{Kocharovskaya1988, Khanin1990, Scully1989, Harris1989, Nottelmann1993, Fry1993, Veer1993, Zibrov1995}. 

The basic idea behind lasing without inversion\cite{Kocharovskaya1992, Scully1992, Scully1994} (LWI)
is similar to conventional lasers -- to prepare a medium where emission is favored 
over absorption, so 
that resonant light is amplified as it propagates in the medium.
The difference lies in the medium preparation.
Instead of creating population inversion between the lasing states, quantum interference is used to suppress photo-absorption while keeping 
photo-emission intact. 
In the simplest scheme, two 
lower-lying states $|g_1\rangle$ and $|g_2\rangle$ are 
each coupled to a common upper state $|e_1\rangle$,
see Fig.1(a).
One can prepare a coherent superposition of the two lower states, such that the transition from this 
superposition to the upper state vanishes due to destructive interference of the transition amplitudes 
$|g_1\rangle\rightarrow |e_1\rangle$   and 
$|g_2\rangle\rightarrow |e_1\rangle$. 
Now, any population pumped incoherently into the 
upper state can provide gain. 

A number of different schemes for lasing without inversion have been developed\cite{Kocharovskaya1992, Scully1992}. Taking advantage of   
destructive interference of different pathways in
absorption, they generally strive 
to maintain a specific phase relationship between 
the lower-lying states, which carry most of the population.
Here we present a scheme that does not follow
this tradition. It uses only the natural 
dynamics of a multi-level quantum system and
requires no coherence between the excited
and the lower electronic states; effectively,
lasing without inversion comes 'for free'.
We also show that this mechanism is active
in the highly efficient generation of 391~nm radiation
during propagation
of intense femtosecond laser pulses in 
air\cite{Luo2003, Yao2011, Yao2013, Liu2013, Chu2013}, under standard conditions where the process known 
as "laser filamentation"\cite{Braun1995, Couairon2007, Berge2007} leads to self-guiding of light. 

Identifying the mechanism responsible for this 
effect, commonly referred to as  
'air lasing', has been a long-standing puzzle\cite{Ni2013, Zhang2013, Li2014, Mitryukovskiy2015, Liu2015, Xu2015, Yao2016, Kartashov2014, Richter2017, Azarm2017, Lei2017, Zhong2017, Xu2017, Liu2017, Britton2018, Mysyrowicz2018, Xu2018, Zhong2018, Arissian2018, Britton2019, Li2019a, Ando2019, Li2019, Zhang2019, Zhang2019a,ZhangQian2019}.
The main difficulty in resolving this puzzle
stems from the apparent lack of a general
physical mechanism capable of generating population inversion for the dominant observed amplification line
around 391~nm, for standard filamentation conditions including the clamping of 
the laser intensity around
$I\sim 10^{14}$~W/cm$^2$. The 391~nm line 
corresponds to the transition
between the two ground vibrational levels
$\nu'', \nu' = 0$ 
of the electronic states
$X^2\Sigma_g^+$ (denoted as $X$) 
and $B^2\Sigma_u^+$ (denoted as $B$)
in N$_2^+$. We believe that 
our inversionless mechanism
provides the key missing component of this puzzle. 
Compared to the recent proposal in Ref.\cite{Mysyrowicz2018}, we fully account for the vibrational and rotational dynamics of the molecule; our lasing mechanism remains operative even in the absence of initial coherence between the ionic states. 

\section*{Results and discussion}
In its simplest form, our amplification mechanism can be referred to as rotational quantum beat lasing. 
It is illustrated in Fig.1(b)-(d), and can be easily understood in the time domain. 
\begin{figure}
 \centering
 \includegraphics[width=0.92\textwidth]{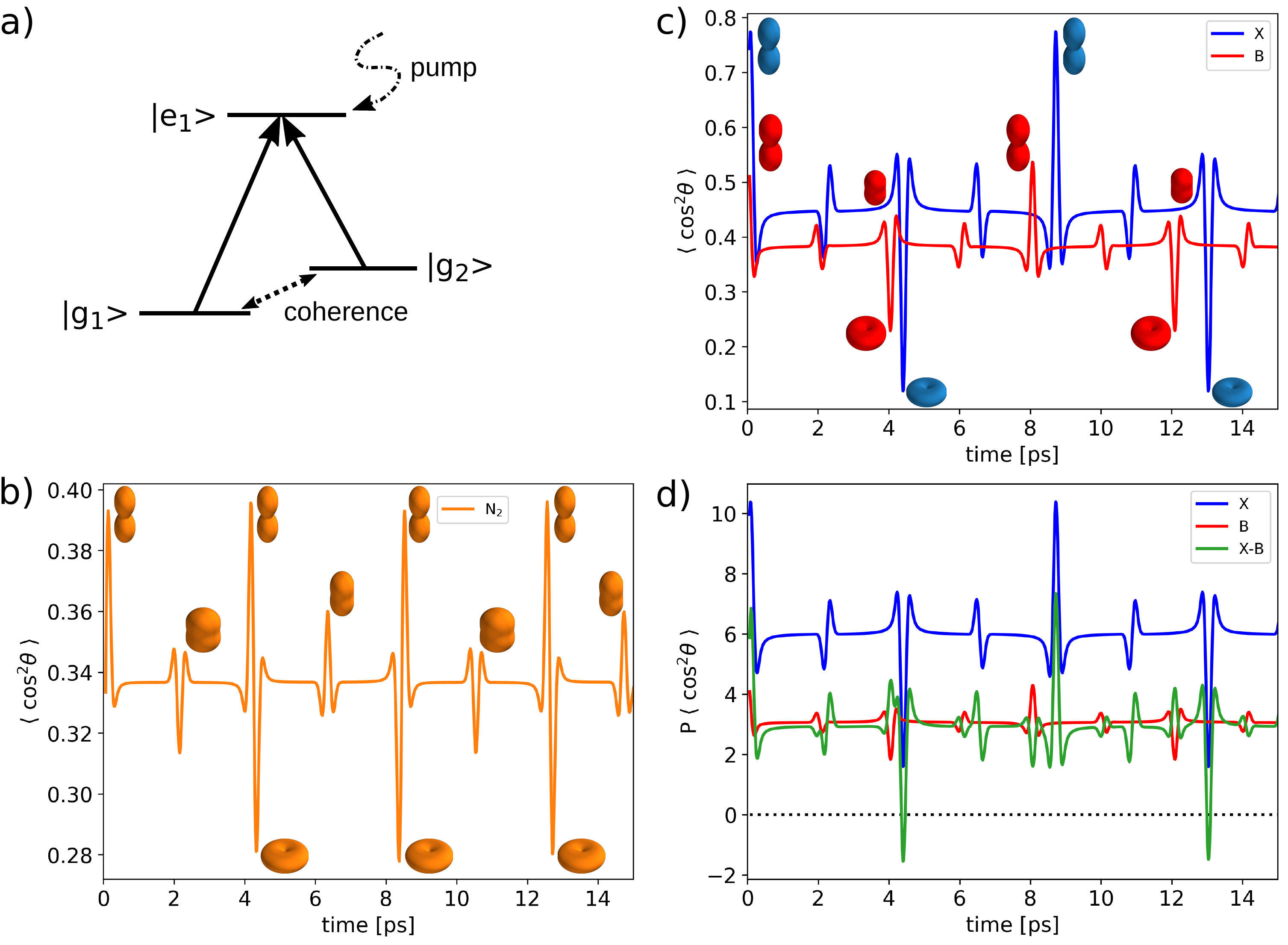}
 \caption{Schematic of a rotational quantum beat laser without inversion in the time domain using the example of N$_2^+$. Calculations are performed for a $23$~fs, $800$~nm, $10^{14}$~W/cm$^2$ pump pulse and room temperature ($298$~K). The 3D shapes in (b,c) sketch the angular distributions of the rotating molecular ensembles.
 (a) Conventional lasing-without-inversion gain medium with low-frequency coherence between a pair of states. 
 (b) Molecular alignment of N$_2$ at $298$~K induced by the intense femtosecond pump pulse.
 (c) Rotational dynamics in the two vibronic states $|X,\nu''=0\rangle$ and $|B,\nu'=0\rangle$ of N$_2^+$ induced by the same pump. (d) Gain windows (at $\tau_1 \simeq 4.3$~ps and $\tau_2\simeq 13.0$~ps) without population inversion at a parallel transition between the two vibronic states are enabled by the different angular distribution 
 geometries of the rotating molecular ensembles in the upper and the lower states; no coherence between the two rotational manifolds is required. 
 }
 \label{fig:energy_level_diagrams}
\end{figure}
When an intense femtosecond laser pulse interacts with a diatomic molecule, or its ion, 
it induces molecular alignment\cite{Seideman1999, Dooley2003}. 
This alignment revives periodically after the pulse is gone, Fig.1(b). 
The revival dynamics demonstrates a sequence of alignment and anti-alignment with the revival period controlled by the rotational constant $B_0$.
Fig.1(b) shows the evolution of 
the characteristic alignment measure, 
$\langle\cos^2\theta\rangle(t)$,
for a nitrogen 
molecule at room temperature (298~K), after interacting with a 23~fs FWHM, 800~nm pump pulse 
with a peak intensity of $10^{14}$~W/cm$^2$ (see Methods).
Its minima correspond to anti-alignment,
its maxima to maximum alignment with the direction 
of the linearly polarized, aligning laser field.

Suppose now that similar rotational
dynamics is induced in two electronic (vibronic) states, 
$|1\rangle$ and $|2\rangle$, with a dipole-allowed 
electronic transition between them.  
In diatomics, electronic transitions can be either parallel or perpendicular to the molecular axis; 
for definiteness, let 
this transition be parallel. 
Its probability is then proportional to
$\cos^2 \theta$, where $\theta$ is the angle between the
molecular axis and the laser electric field 
inducing the transition.
For a molecule rotating in the lower vibronic state, the 
probability of parallel absorption 
$|1\rangle\rightarrow|2\rangle$ 
thus depends on $\langle\cos^2\theta\rangle_{\rm 1}(t)$ 
averaged over the rotational dynamics in state $|1\rangle$, which is controlled by its rotational constant $B^1_0$.
Conversely, for a molecule rotating in the upper vibronic state $|2\rangle$, the 
emission probability $|2\rangle\rightarrow|1\rangle$ 
depends on $\langle\cos^2\theta\rangle_{\rm 2}(t)$, 
with its time-dependence controlled by $B^2_0$.

Opportunities for inversionless 
amplification  of a short probe pulse,
polarized parallel to the pump 
and delayed by $\tau$,
arise when the $|1\rangle$ state molecules are anti-aligned,
so that $\langle\cos^2\theta\rangle_{\rm 1}(\tau)$ is minimum and the parallel $|1\rangle \rightarrow |2\rangle$ absorption is suppressed.  
This opportunity is further enhanced if 
the $|2\rangle$ state molecules are aligned at this time,
so that $\langle\cos^2\theta\rangle_{\rm 2}(\tau)$ is maximum and the parallel $|2\rangle \rightarrow |1\rangle$ emission is enhanced. 
This possibility to use molecular alignment for inversionless amplification was previously pointed out by A. K. Popov and V. V. Slabko\cite{Popov2005}. 
The converse is true if the $|1\rangle$ state molecules are aligned when the probe pulse arrives, while
the $|2\rangle$ state molecules are anti-aligned, enhancing absorption.
As long as $B^1_0$ and $B^2_0$ are different,
the two rotations will go out of sync, arriving at the point where the lower state is anti-aligned and the upper is still aligned and vice versa (Fig 1c). 
Overall, temporal windows of gain will be followed by 
windows of loss, 
leading to the rotational quantum beats 
in the time-resolved gain-loss of the short probe pulse. 

Figs.1(c,d) give the specific example. Fig.1(c) shows the rotational dynamics in the two ground vibrational levels
$\nu'', \nu' = 0$ 
of the electronic states
$X$ and $B$ in N$_2^+$, 
induced by the same pump pulse interacting with N$_2$, Fig.1(b).
The short pump pulse impulsively aligns
the neutral N$_2$ molecules, 
generating the rotational dynamics in Fig.1(b).
It also 
ionizes some of the N$_2$ molecules, 
producing molecular ions predominantly 
in the ground $X$, but also in the excited
$A$ ($A^2\Pi_u$) and $B$ states, and
continues to align them. 
After the end of the pulse, the ions
continue to rotate, 
reaching maximum alignment at $t\simeq 90$~fs for the $X$ and $B$ states, 
see Fig.1(c), followed by periodic 
revivals of alignment and anti-alignment. 
They correspond to maxima and minima in the 
rotational ensemble-averaged 
$\langle\cos^2\theta\rangle_{\rm X,B}(t)$, where 
subscripts denote the ionic states (see Methods).
These revivals are different due to slight differences in the
rotational constants $B^{\rm X,B}_0$.

Frequency-integrated absorption at the parallel $|X, \nu''=0\rangle \rightarrow |B, \nu'=0\rangle$ transition
behaves as $W_{\rm abs}(\tau)\propto 
P_{X}\langle\cos^2 \theta\rangle_{X}(\tau)$, where
 $P_{X}$ is the $|X,\nu''=0\rangle$-state 
population (see Methods).  
In emission,  
$W_{\rm e}(\tau)\propto P_{B}\langle\cos^2 \theta\rangle_{B}(\tau)$. 
Gain windows open 
when $P_{B}\langle\cos^2 \theta \rangle_{B}(\tau)>
P_{X}\langle\cos^2 \theta \rangle_{X}(\tau)$, see Fig.1(d).
The populations of the $X$ and $B$ states, $P_{\rm X,B}$,
determine the length and the strength of the
gain windows, however, overall electronic population inversion
is not needed, Fig.1(d).
Comprehensive quantitative analysis 
presented below shows  that
this naturally arising amplification mechanism  does indeed
lead to gain 
at the rotational frequency band around 391~nm
during filamentation of
intense femtosecond laser pulses in air or N$_2$ gas,
including the intensity regime 
$I \lesssim 10^{14}$~W/cm$^2$
pertinent to laser filamentation.

Most experiments studying this effect\cite{Ni2013, Yao2013, Zhang2013, Li2014, Kartashov2014, Liu2015, Yao2016, Lei2017, Zhong2017, Xu2017, Zhong2018, Britton2018, Mysyrowicz2018, Xu2018, Arissian2018, Li2019, Britton2019, Ando2019, Zhang2019, Zhang2019a} 
are done in a 
pump-probe scenario: an intense pump pulse
prepares the gas and transient absorption (or gain)
of a delayed probe pulse is measured. The 
theoretical results presented below focus on this scenario,
with the pump pulse carried at 800 nm.
Details are given in the Methods section, here 
we outline key issues requiring particular attention
when addressing this rather challenging problem quantitatively. 

Our theoretical description accounts for 
laser-induced alignment of the neutral N$_2$ molecule, 
its alignment-dependent 
strong-field ionization into the  laser-dressed 
states of  the molecular ion, and full laser-induced electronic, vibrational and rotational dynamics in the ion 
involving the $X, A, B$ states. 
Our ab-initio simulations of strong-field ionization use 
the method of Ref.\cite{Spanner2009}, and allow us to
evaluate the excitation of the different ionic states induced by the recollision of the photoelectron with the parent ion in the $X$ state\cite{Mitryukovskiy2015, Liu2015, Liu2017, Mysyrowicz2018, Britton2018, Li2019a}. 
The role of recollision is gauged by absorbing the 
electron wavepacket before it is turned around towards 
the parent ion. Eliminating the 
recollision reduces the population of the $B$ state 
by about 1\%,
compared to the case when the recollison is included.
Thus, the recollision can be neglected and
the photo-electron can be integrated out. Here,  
one must account for the electron-ion entanglement. 
Since the $X$ and $B$ states have opposite parity, the  
photo-electron wavepackets correlated to them will also carry opposite parity, remaining orthogonal to each 
other. This eliminates the coherence between the $X$ and $B$ states that could have been produced during strong-field ionization.

Regarding optical $X\rightarrow B$ excitation by an 
800~nm pump, the $|X, \nu''=0\rangle \rightarrow |B, \nu'=0\rangle$ transition corresponds to absorption of two photons, 
which is parity forbidden; noticeable excitation is only generated at very high intensities, $I\gtrsim 4 \times 10^{14}$
W/cm$^2$ (the domain of several recent pump-probe experiments but not relevant for standard laser filamentation conditions), when the system is strongly distorted and higher-order multiphoton transitions can occur. In this context, note that strong-field ionization
populates not field-free\cite{Xu2015, Li2019, Ando2019, Zhang2019b,ZhangQian2019} but already polarized (dressed by the field) ionic $X$, $A$, and $B$ states. Indeed optical tunneling results from polarization of the many-body wave function of the neutral as one of the polarized electrons leaks through the potential barrier, leaving other electrons polarized.
Comparison to ab-initio simulations\cite{Spanner2009} show that initializing population in the field-free ionic states at the peaks of the instantaneous electric field, where
strong-field ionization takes place, generates spurious excitations of the $A$ and $B$ states due to effectively abrupt turn-on of the laser-ion interaction.

Moreover, the initial thermal rotational distribution
must be included. Each rotational state in the initial
thermal distribution gives rise to a rotational
wavepacket induced by the pump. Each wavepacket
leads to coherent effects in emission and absorption; 
these contributions 
are added incoherently
with the corresponding weights.

With these aspects accounted for, our calculations reveal a
robust physical mechanism leading to gain at the 391 nm line.
In contrast to Ref.\cite{Mysyrowicz2018}, 
it does not require the coherence between the $X$ and $B$ states, and is based on the mechanism described in Fig.1.

In the frequency domain, the rotational quantum beat mechanism can be understood as follows. 
The short, intense pump pulse aligning and ionizing the N$_2$ molecules, generates broad rotational distributions in the $X$ and $B$ states
of N$_2^+$, with coherent population of many adjacent rotational states with similar amplitudes in each electronic state. 
Fig.2(a) shows the rotational distributions in $|X,\nu''=0\rangle$ and $|B,\nu'=0\rangle$ for the case of the 800~nm, 23~fs FWHM pump pulse with an intensity of $I=10^{14}$~W/cm$^2$ and the N$_2$ molecules initially at room temperature (298~K). 
\begin{figure}
 \centering
 \includegraphics[width=0.81\textwidth]{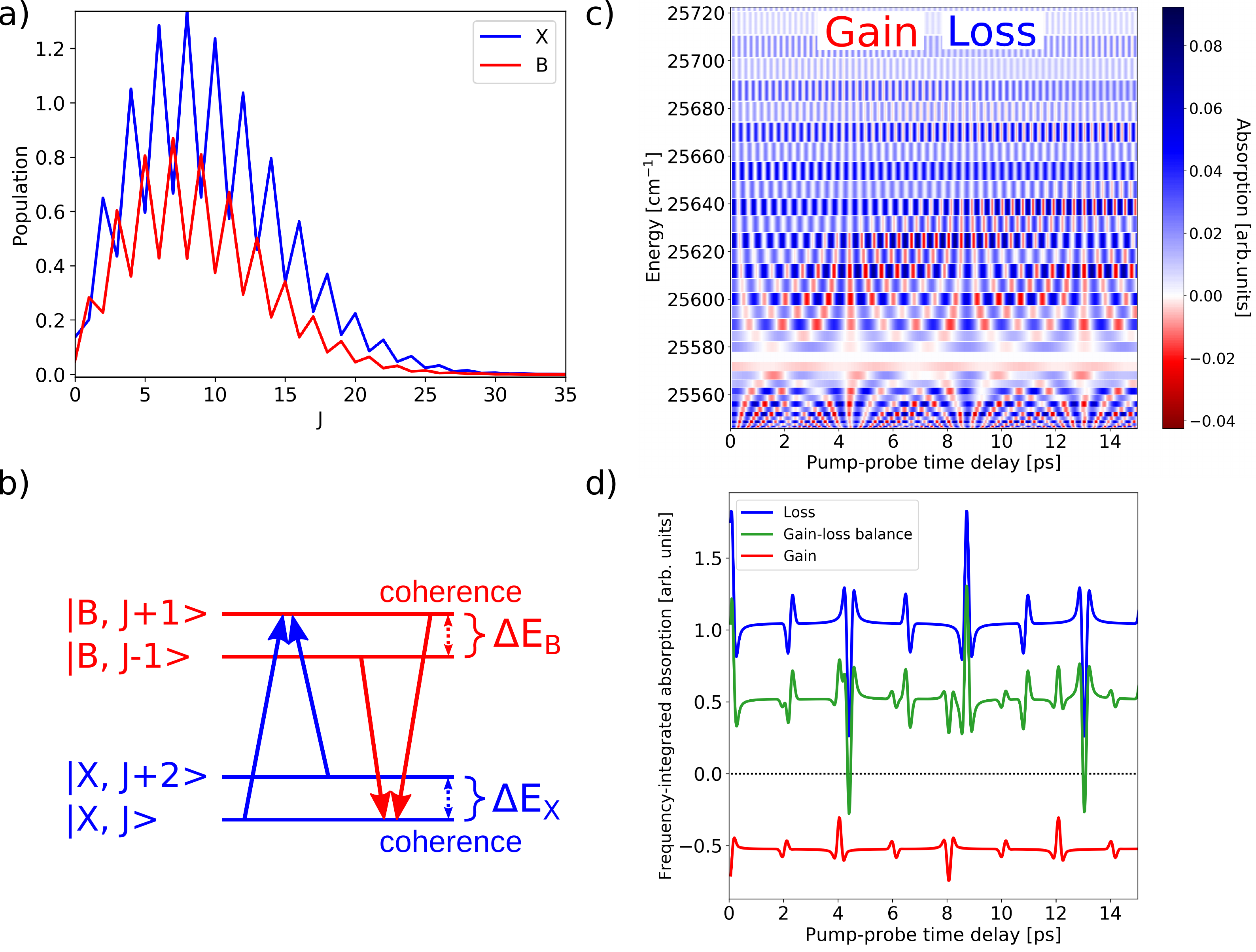}
\caption{Rotational quantum beat lasing without inversion in the frequency domain using the example of N$_2^+$. Calculations are performed for the same pump pulse and temperature as in Fig.1 and a $20$~fs, $391$~nm, $10^{11}$~W/cm$^2$ probe pulse. 
(a) Rotational distributions in the $X$ and $B$ states of N$_2^+$ formed at the end of the interaction with the intense femtosecond pump pulse. (b) Four-level system controlling absorption and emission at the $|X,J\rangle \rightarrow |B,J+1\rangle$ transition. (c) Full simulation of the frequency-resolved gain (red) and loss (blue) of the weak probe interacting with the pumped system as a function of pump-probe delay.  
(d) Absorption (blue) and emission (negative absorption) (red) integrated over all frequencies follow the alignment dynamics in Fig.1(c,d). The total gain-loss balance integrated over all frequencies (green) follows the pattern of $P_X\langle \cos^2 \theta \rangle_X - P_B\langle \cos^2 \theta \rangle_B$ in Fig.1(d), showing the same gain windows for the inversionless medium ($P_X>P_B$).
 }
 \label{fig:absorption_emission_signal}
\end{figure}

Consider now frequency-resolved transient absorption 
of a short probe pulse at the dipole-allowed
transition $|X,J\rangle \rightarrow |B,J+1\rangle$ 
(with the vibrational quantum number equal to zero
in both cases). 
For each initial rotational state of the neutral,
rotational coherence generated in the 
$X$ state of the ion
means that the absorption $|X,J\rangle \rightarrow |B,J+1\rangle$, 
stimulated by a broadband probe, 
is inevitably accompanied by the absorption
$|X,J+2\rangle \rightarrow |B,J+1\rangle$ stimulated
by the same probe, Fig.2(b).
Their interference is governed by 
the relative phase between the two lower states,
$\phi_{X}(J+2,J)(t)=\Delta E_{ X}(J+2,J)t+\phi^{(0)}_{ X}(J,J+2)$. 
Here
$\Delta E_{X}(J+2,J)=B^{X}_0 \left[(J+2)(J+3)-J(J+1) \right]$ is the distance between the two states in $X$ and $\phi^{(0)}_{X}$ is their relative phase at the start of the field-free evolution when the pump is over.
Destructive interference between the 
$|X,J\rangle \rightarrow |B,J+1\rangle$
and $|X,J+2\rangle \rightarrow |B,J+1\rangle$ transitions
occurs when $\phi_{X}(J+2,J)(t)=(2N\pm 1)\pi$ ($N$ is integer), resulting in 
suppression of absorption of a short probe pulse arriving at this moment.
Conversely, constructive interference enhances absorption when 
$\phi_{X}(J+2,J)(t)=2N\pi$. 
This leads to the quantum beat in the time-dependent absorption rate of the short probe pulse with frequency $\Delta E_{X}(J+2,J)$.

Emission at the same frequency, $|B,J+1\rangle\rightarrow |X,J\rangle$, is inevitably accompanied by the transition 
$|B,J-1\rangle \rightarrow |X,J\rangle$, Fig.2(b).
Their interference is governed by 
the relative phase 
$\phi_{B}(J+1,J-1)(t)=\Delta E_{B}(J+1,J-1)t + \phi^{(0)}_{B}(J-1,J+1)$ with $\Delta E_{B}(J+1,J-1)=B^{B}_0 \left[(J+1)(J+2)-(J-1)J \right]$.
Constructive interference between 
$|B,J+1\rangle \rightarrow |X,J\rangle$ and
$|B,J-1\rangle \rightarrow |X,J\rangle$ in 
emission occurs when $\phi_{B}(J+1,J-1)(t)=2N\pi$,
while destructive interference occurs when 
$\phi_{B}(J+1,J-1)(t)=(2N\pm 1)\pi$, leading to the quantum beat in the time-dependent emission rate with frequency $\Delta E_{B}(J+1,J-1)$.

The difference between the rotational constants
$B^{X}_0$ and $B^{B}_0$ means that there are 
time-windows where destructive interference of
the $|X,J\rangle \rightarrow |B,J+1\rangle$ and
$|X,J+2\rangle \rightarrow |B,J+1\rangle$ transitions in 
absorption (when $\phi_{X}(J+2,J)(t)=(2N\pm 1)\pi$)
coincides with constructive interference of the
$|B,J+1\rangle \rightarrow |X,J\rangle$
and $|B,J-1\rangle \rightarrow |X,J\rangle$ transitions 
in emission (when
$\phi_{B}(J+1,J-1)(t)=2N\pi$). 
This leads to frequency-resolved ($J$-dependent) gain and 
loss windows as a function of the pump-probe 
time-delay, shown in Fig.2(c) for a weak 
20~fs Gaussian probe pulse 
with a central wavelength of 
391~nm.

As many four-level systems 
such as the one shown in Fig.2(b) 
are formed by the pump, they 
generate a rich gain (red)-loss (blue) pattern 
of frequency-resolved transient absorption of the 
probe pulse, see Fig.2(c). 
Each vertical line shows the frequency-resolved absorption 
for a particular pump-probe time delay, 
integrated over the full duration of the probe pulse.
The horizontal white line around $\tilde{\nu}\simeq 25575$~cm$^{-1}$ 
marks the energy spacing $\tilde{\nu}_{00}$ 
between the two ground ro-vibrational levels of the electronic states $X$ and $B$.
The spectral lines above this energy are associated 
with the higher-energy branch (R branch).
Those below 25575 cm$^{-1}$ to the lower energy branch (P branch).

Time-dependent gain emerges for virtually all 
pump-probe time delays
and many transition frequencies, 
especially those associated with the lower-energy P-branch. 
Most notably, for those time delays at which the $X$ state molecules are anti-aligned with the probe polarization direction, e.g. around $\tau \simeq 4.3$~ps, 
amplification occurs across the whole range of frequencies.
Fig.2(c) incorporates the 
comprehensive modeling of the whole process, starting with 
the thermal ensemble of neutral molecules 
and accounting for 
(i) alignment of the neutral molecule, (ii) its angle-dependent
strong-field ionization into the laser-dressed ionic states,
with the ratio of $B$ to $X$ state populations at $20\%$ upon ionization,
(iii) entanglement between the photo-electron and the ion,
which negates the $X$-$B$ coherence once the 
photo-electron is integrated out, 
(iv) coupled electronic, vibrational, and rotational
dynamics in the ion, and (v) 
comprehensive calculations of transient absorption and emission of the probe pulse including all interferences. 
Note that gain windows emerge
with neither electronic nor rotational population 
inversion, Fig.2(a). 

Fig.2(d) connects the frequency-domain picture of Fig.2(c) 
to the time-domain picture of Figs.1(c,d).
It shows the frequency-integrated 
absorption from the $X$ state (blue line) and  
emission (negative absorption) from the $B$ state (red line) stimulated by the short 
probe pulse, as a function of the pump-probe delay. 
The pattern of each line
follows the alignment measure of the respective ion in Fig.1(c). This is not a coincidence.
The same phase differences that control the interferences 
in transient  absorption by each ($|X,J\rangle, |X,J+2\rangle$) 
pair, control their contribution to 
the ensemble-averaged $\langle \cos^2\theta\rangle_{\rm X}$ 
measure of the field-free rotations in the $X$ state.
The same applies to emission by each ($|B,J-1\rangle, |B,J+1\rangle$) 
pair and the ensemble-averaged $\langle \cos^2\theta\rangle_{\rm B}$ 
measure.
As shown in Methods, integrating the frequency-resolved 
absorption over all frequencies yields the total transient absorption probability $W_{\rm abs}(\tau)\propto 
P_{\rm X}\langle\cos^2(\theta)\rangle_{\rm X}(\tau)$.
Analogous results hold for transient emission. 

The green line in Fig.2(d) shows the overall gain-loss balance, integrated over all energies. 
As expected, the strongest gain windows coincide with anti-alignment of the $X$ state, cf. Fig.1(c,d), when the  gain-loss pattern in Fig.2(c) shows gain across the whole spectrum.
If the $X$ and $B$ state molecules would anti-align simultaneously during their field-free evolution, 
the net gain would vanish.

Results presented in Fig.2 are robust with respect to the pump parameters, and the lasing mechanism described above is completely general. 
The only necessary ingredients are non-negligible population in the $B$ state and molecular rotations, both unavoidable during the interaction with an intense laser pulse.
For the pump pulse parameters and temperature used for Figs.1,2, our simulations show that net gain sets in already at the ratio $P_B/P_X \simeq 10\%$ upon ionization.

Fig.3 shows results for two other pump intensities
realistic in laser filamentation.
\begin{figure}
 \centering
 \includegraphics[width=\textwidth]{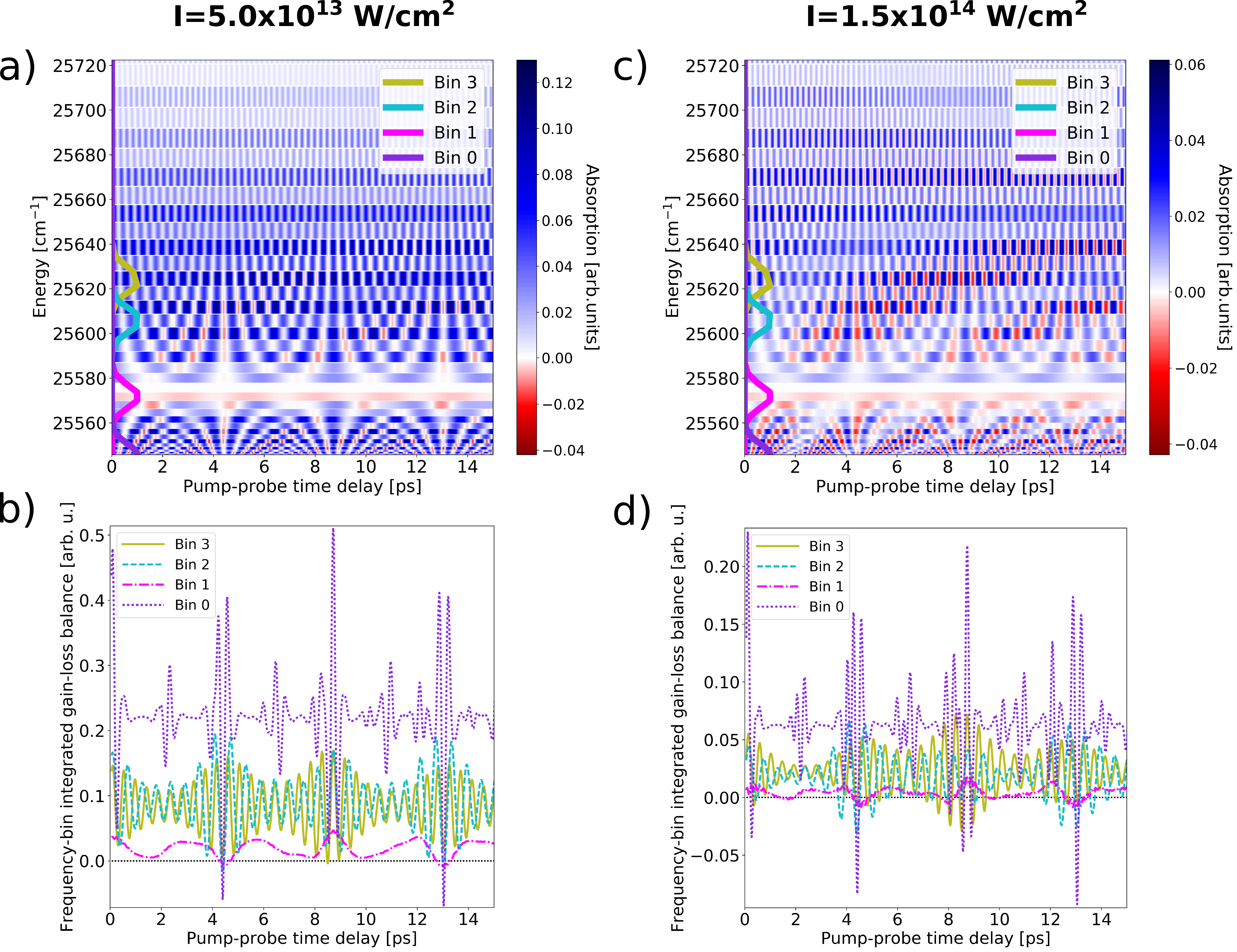}
\caption{Rotational quantum beat lasing in N$^+_2$ for different intensities realistic in femtosecond laser filamentation experiments. Calculations are performed for a $23$~fs, $800$~nm pump pulse and the same probe pulse and temperature as in Fig.2.  
(a,c) Same as Fig.2(c), but for a pump intensity of (a) $I=5.0\times10^{13}$~W/cm$^2$ and (c) $I=1.5\times10^{14}$~W/cm$^2$.
(b,d) Gain-loss balance integrated over the frequency windows (bins) indicated in panels (a,c). 
For both intensities, gain windows emerge without population inversion of the medium ($P_X>P_B$) (Gain windows also emerge for the total gain-loss balance integrated over all frequencies, analog to Fig.2d).
}
 \label{fig:absorption_emission_signal_diff_int}
\end{figure}
The expected "clamping" intensity is $I\lesssim 1\times 10^{14}$~W/cm$^2$, but pulse compression during
filamentation suggests that short intensity spikes can 
reach\cite{Gaarde2009,Mitryukovskiy2015clamping} $I\sim 2\times 10^{14}$~W/cm$^2$. 
For $800$~nm driver fields with intensities $I\lesssim 2\times10^{14}$~W/cm$^2$, the $B$ state population is not strongly affected by laser-driven excitations from $X$.
However, the higher the intensity, the stronger the depletion of the $X$ state into $A$\cite{Xu2015, Yao2016} (especially for pump pulses shorter or comparable to the $\sim18$~fs vibrational period in the $A$ state), yielding an overall change of the gain-loss balance in favor of gain and hence the emergence of additional gain windows in time, Figs.3(d).

The results presented in Fig.3 are in remarkably good agreement with recent experimental findings\cite{Arissian2018,Britton2019}. 
Ref.\cite{Arissian2018} reports on pump-probe femtosecond filamentation experiments in a nitrogen gas cell, measuring the delay-dependent amplification of the probe around the $391$~nm line using a high-resolution spectrometer. 
Our simulations closely reproduce all the main features of the measured time- and frequency-resolved gain, including the  parabolic structure of the gain, the fast temporal modulation of each rotational line that increases with $J$ in the R branch, and the slower temporal modulation, reflecting the molecular alignment dynamics and giving rise to the distinct gain window at $\tau \sim4.3$~fs where amplification occurs across the whole spectrum. Experiments that do not resolve each individual rotational line observe (the weighted sum of) frequency window-integrated, highly oscillating gain-loss structures, such as the ones shown in Figs.3(b,d). For increasing pump intensity, the oscillating gain starts to occur for virtually all pump-probe time delays, as observed in the nitrogen gas jet experiments in Ref.\cite{Britton2019}.

In the present mechanism, the gain window travels 
with the group velocity of the pump pulse, 
so that amplification 
occurs only in the forward direction. To 
achieve lasing in the backward direction,  
highly desirable for remote sensing, 
one would greatly benefit from creating 
population inversion, 
at least between the rotational states. To this end, 
a sequence
of two well-timed pump pulses can be used to
control rotations of $X$ and $B$ states of the ion.
Their different rotational periods mean that
the second pulse can be timed to simultaneously accelerate
rotations of the $B$-state and slow down the 
rotations of the $X$ state, generating
rotational inversion. Optimization
of such pulse sequence, including their carrier 
frequencies, energies, and time-delay, should
open a route for achieving robust and strong 
backward lasing in open air.

\section*{Acknowledgements}
We acknowledge many fruitful and stimulating discussions with Daniil Kartashov, Andrius Baltuška, Mathew Britton and Paul Corkum.
M.I. acknowledges support from the Deutsche For-
schungsgemeinschaft (DFG) Quantum Dynamics in Tailored Intense Fields (QUTIF) grant no. IV 152/6-1, and from the Engineering and Physical Sciences Research Council/Defence Science and Technology
Laboratory (EPSRC/DSTL) Multidisciplinary University Research Initiative (MURI) grant no. EP/N018680/1.
O.S. acknowledges support from the DFG Schwerpunktprogramm 1840 Quantum Dynamics in Tailored
Intense Fields project SM 292/5-1, and Molecular Electron Dynamics Investigated by Intense Fields and from the
Attosecond Pulses (MEDEA) project, which has received funding from the European Unions Horizon 2020
research and innovation programme under the Marie Sklodowska-Curie grant agreement no. 641789.
S.H. acknowledges support from the Laboratoire d'Excellence Physique: Atomes Lumière Matière (LabEx PALM) overseen by the Agence Nationale pour la Recherche as part of the Investissements d'Avenir program (ANR-10-LABX-0039).

% \section*{References}
\bibliography{references.bib}

\end{document}